\documentclass[aps,pre,reprint,amsmath,amssymb,longbibliography,lengthcheck,showpacs,superscriptaddress,floatfix]{revtex4-2}
\usepackage{enumerate}
\usepackage{graphicx}
\usepackage{color}
\begin{document}
%
\title{Universal mechanism of shear thinning in supercooled liquids}
%
\author{Hideyuki Mizuno}
\email{hideyuki.mizuno@phys.c.u-tokyo.ac.jp}
\affiliation{Graduate School of Arts and Sciences, The University of Tokyo, Tokyo 153-8902, Japan}
\author{Atsushi Ikeda}
\affiliation{Graduate School of Arts and Sciences, The University of Tokyo, Tokyo 153-8902, Japan}
\affiliation{Research Center for Complex Systems Biology, Universal Biology Institute, The University of Tokyo, Tokyo 153-8902, Japan}
\author{Takeshi Kawasaki}
\affiliation{Department of Physics, Nagoya University, Nagoya 464-8602, Japan}
\author{Kunimasa Miyazaki}
\affiliation{Department of Physics, Nagoya University, Nagoya 464-8602, Japan}
%
\date{\today}
%
\begin{abstract}
Soft glassy materials experience a significant reduction in viscosity $\eta$ when subjected to shear flow, known as shear thinning.
This phenomenon is characterized by a power-law scaling of $\eta$ with the shear rate $\dot{\gamma}$, $\eta \propto \dot{\gamma}^{-\nu}$, where the exponent $\nu$ is typically around $0.7$ to $0.8$ across different materials.
Two decades ago, the mode coupling theory (MCT) suggested that shear thinning occurs due to the advection.
However, it predicts too large $\nu = 1~(> 0.7$ to $0.8)$ and overestimates the onset shear rate by orders of magnitude.
Recently, it was claimed that a minute distortion of the particle configuration is responsible for shear thinning.
Here we extend the MCT to include the distortion, and find that both advection and distortion contribute to shear thinning, but the latter is dominant.
Our formulation works quantitatively for several different glass formers.
We explain why shear thinning is universal for many glassy materials.
\end{abstract}
%
\maketitle
%
\section*{Introduction}
Nonlinear rheology is commonly observed in complex fluids and soft materials~\cite{Rheology}.
In particular, supercooled glassy liquids exhibit a significant reduction in viscosity $\eta$ and relaxation time $\tau_\alpha$ when subjected to fast shear flow, a phenomenon known as shear thinning.
Both experiments~\cite{Webb_1998,Kato_1998,Bessel_2007} and simulations~\cite{Yamamoto_1998,Berthier_2002,Furukawa_2009,Mizuno_2012} have shown that $\eta$ and $\tau_\alpha$ follow a power-law scaling with the shear rate $\dot{\gamma}$ as $\eta,\ \tau_\alpha \propto \dot{\gamma}^{-\nu}$, where the exponent $\nu$ remains consistent across different materials, typically around $0.7$ to $0.8$.
Understanding shear thinning is crucial not only for the manufacturing and processing of materials but also for broader physical phenomena such as volcanoes and earthquakes.
However, we have yet to understand the mechanism responsible for this nonlinear flow and its universal nature.
Various theories have been proposed in the past, including the soft glassy rheology theory~\cite{Sollich_1997}, the shear transformation zone theory~\cite{Langer_2008}, and the elastoplastic model~\cite{Nicolas_2018}.

Among these theories, the mode coupling theory~(MCT) is a first-principles theory, which was originally developed to explain the equilibrium dynamics of supercooled liquids near the glass transition point~\cite{MCT}.
The theory describes the slow glassy dynamics in terms of the caging effect; particles are trapped in the cages formed by their neighbors until the structure undergoes reconfiguration at the equilibrium relaxation time $\tau_{\alpha 0}$, which diverges at the dynamical transition point in the mean-field limit.
Note that we denote equilibrium values by the subscript 0 throughout this article.
In later years, the MCT has been generalized to sheared liquids~\cite{Fuchs_2002,Miyazaki_2002}. 
The sheared MCT explains that advection induced by the shear flow breaks the cages and accelerates the dynamics.
Shear thinning begins when the timescale of shear, $\dot{\gamma}^{-1}$, becomes comparable to $\tau_{\alpha 0}$, that is, the onset shear rate is $\dot{\gamma}_c \sim \tau_{\alpha 0}^{-1}$. 
The theory then predicts that as the shear rate further increases, $\tau_\alpha$ decreases as $\dot{\gamma}^{-1}$ and thus the thinning exponent $\nu=1$.

Although the sheared MCT provides a qualitative explanation for the reduction in relaxation time and viscosity, its prediction of $\nu = 1$ is larger than the observation of $\nu \sim 0.7$ to $0.8$ in many previous works~\cite{Yamamoto_1998,Berthier_2002,Furukawa_2009,Mizuno_2012,Webb_1998,Kato_1998,Bessel_2007}.
Moreover, the theory overestimates the values of the onset shear rate $\dot{\gamma}_c$, which have been found to be orders of magnitude smaller than the theoretical prediction $\tau_{\alpha 0}^{-1}$, {\it i.e.}, $\dot{\gamma}_c \ll \tau_{\alpha 0}$, in experiments and simulations~\cite{Furukawa_2009,Webb_1998,Miyazaki_2006,Lubchenko_2009}.
These discrepancies between the theory and the observations have remained unaddressed for more than two decades.

Recently, Furukawa~\cite{Furukawa_2017,Furukawa2_2023,Furukawa_2023} has proposed a semi-microscopic theory to explain the shear thinning in supercooled liquids, which is distinct from the advection scenario of the sheared MCT.
The theory claims that anisotropic distortion of the particles' configuration due to shear flow, rather than advection, is responsible for shear thinning.
Although this distortion is tiny in dense glassy fluids~\cite{Ronis_1984,Hanley_1987,Iwashita_2012,Yamaguchi_2018}, it reduces effective density for fragile glass formers~\cite{Furukawa_2017,Furukawa2_2023} or effective activation energy for strong glass formers~\cite{Furukawa_2023}, which, in turn, induces a drastic acceleration of the dynamics and causes the shear thinning.
Based on this distortion scenario, Furukawa succeeded in quantitatively explaining the observed small thinning exponent, $\nu <1$, and the small onset rate, $\dot{\gamma}_c \ll \tau_{\alpha 0}^{-1}$, for both fragile and strong glass formers.

Questions naturally arise within us.
(i) Does the advection scenario of the sheared MCT fail to explain the shear thinning in supercooled liquids?
(ii) Does the distortion scenario work universally in different types of supercooled liquids?
If so, (iii) can we renovate the sheared MCT by integrating the distortion effect into the theory and reconciling the theory with the observations?

To answer these three questions, we must first assess the validity of the advection scenario of the sheared MCT.
To do this, we need a model system that can serve as an ideal fluid for testing the mean-field theory of the glass transition.
The Gaussian core model (GCM) is a promising candidate because its slow dynamics are better described by the equilibrium MCT than any other glass-forming liquid~\cite{Ikeda_2011,Ikeda1_2011,Ikeda2_2011,Coslovich_2016}.
Firstly, the GCM is a clean, glassy model that does not require size dispersity~\cite{Ikeda_2011,Ikeda1_2011,Ikeda2_2011}.
The monatomic GCM exhibits slow glassy dynamics close to the dynamical transition point $T_c$ without being affected by unwanted crystallization. 
Secondly, the equilibrium relaxation time of the GCM follows the MCT power-law scaling,
\begin{equation}
\tau_{\alpha 0}(T) \propto (T-T_c)^{-\gamma},
\label{MCTpower1}
\end{equation}
over a wider temperature window than other models of glass formers.
The agreement of the exponent $\gamma \simeq 2.7$ with the MCT is quantitative.
Even the transition temperature $T_c$, routinely used as a fitting parameter, agrees quantitatively with the MCT prediction.
Thirdly, the violation of the Stokes-Einstein~(SE) law is very weak, and the diffusion constant $D_0$ is proportional to $\tau_{\alpha 0}^{-1}$, which is again consistent with the MCT prediction.
Lastly, although a dramatic increase in dynamic heterogeneities accompanies the slow dynamics, the statistics of particles' displacements remain nearly Gaussian~\cite{Coslovich_2016}, and the growth of the dynamical heterogeneities is explained by the inhomogeneous MCT~\cite{Biroli_2006}.
This is in stark contrast with other glass formers, where the separation of fast- and slow-moving clusters of particles characterizes the dynamical heterogeneities~\cite{dhbook}.
Therefore, if the sheared MCT has any prediction regarding the shear thinning, the GCM should be the first model to be compared with that.

In addition to the GCM, we investigate canonical glass formers such as the Kob-Andersen~(KA) model~\cite{Kob_1994}, the soft sphere~(SS) model~\cite{Bernu_1985}, and the van Beest-Kramer-van Santen~(BKS) model~\cite{Beest_1955}.
The KA and SS models are typical fragile glass formers, while the BKS model mimics the silica melt, a representative strong glass former.
We find that the GCM and these different types of supercooled liquids share similar scaling laws in $\tau_\alpha,\ \eta \propto \dot{\gamma}^{-\nu}$ with $\nu \sim 0.7~(< 1)$ and $\dot{\gamma}_c \propto \tau_{\alpha 0}^{-\delta} \ll \tau_{\alpha 0}$ with $\delta \sim 1.4~(> 1)$.
This result indicates that the mechanism of shear thinning is universal, and it can not be explained by the advection scenario of the sheared MCT.

In particular, the GCM does not adhere to the advection scenario, which dictates that the current sheared MCT fails to explain the shear thinning.
We resolve this conundrum by incorporating the distortion effect into the diverging relaxation time and viscosity that the MCT prescribes.
Our analysis of the resulting equation reveals that the thinning exponents of $\nu$ and $\delta$ can be formulated by simple equations, which produce values of $\nu \sim 0.7$ and $\delta \sim 1.4$.
We also extend the schematic model of the sheared MCT, proposed by Fuchs and Cates~\cite{Fuchs_2003}, to account for the distortion effect.
Our theoretical and numerical results resolve long-standing inconsistencies between the theory and the observations in experiments and simulations, and establish a universal mechanism of the shear thinning in supercooled liquids.

\begin{figure}[t]
\centering
\includegraphics[width=0.45\textwidth]{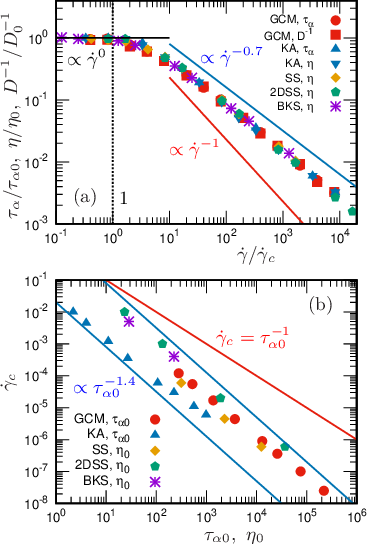}
\caption{\label{fig_data}
{Shear thinning in several different supercooled liquids.}
We present data for the Gaussian core model~(GCM), the Kob-Andersen~(KA) model, the soft sphere~(SS) model, the two-dimensional SS~(2DSS) model, and the van Beest-Kramer-van Santen~(BKS) model.
(a) Plots of $\tau_\alpha/\tau_{\alpha 0}$, $\eta/\eta_0$, or $D^{-1}/D^{-1}_0$ are shown as a function of $\dot{\gamma}/\dot{\gamma}_c$.
The temperature is $T=2.9 \times 10^{-6}$~(GCM), $0.45$~(KA), $0.275$~(SS), $0.577$~(2DSS), and $0.511$~(BKS), all above the dynamical transition temperature $T_c$.
The black line represents $\tau_\alpha/\tau_{\alpha 0},\ \eta/\eta_0,\ D^{-1}/D^{-1}_0 = 1$.
The blue line represents $\propto \dot{\gamma}^{-0.7}$, while the red line refers to $\propto \dot{\gamma}^{-1}$~(advection scenario).
The vertical dotted line indicates the onset shear rate $\dot{\gamma}_{c}$.
(b) $\dot{\gamma}_{c}$ is plotted against $\tau_{\alpha 0}$ or $\eta_0$.
The blue line presents $\dot{\gamma}_{c} \propto \tau_{\alpha 0}^{-1.4}$ or $\eta_0^{-1.4}$, while the red line refers to $\dot{\gamma}_{c} = \tau_{\alpha 0}^{-1}$ or $\eta_0^{-1}$~(advection scenario).
The data for the SS model are extracted from Ref.~\cite{Furukawa2_2023}, the 2DSS model from Ref.~\cite{Furukawa_2017}, and the BKS model from Ref.~\cite{Furukawa_2023}.
}
\end{figure}

\section*{Results and Discussion}
%
\subsection*{Numerical observations}
We perform molecular dynamics~(MD) simulations on the GCM which is subjected to shear flow in three spatial dimensions.
The density is fixed at $\rho=2.0$ where the dynamical transition point has been estimated as $T_c \simeq 2.68 \times 10^{-6}$~\cite{Ikeda_2011,Ikeda2_2011}.
We study a range of temperatures $T$ near $T_c$, so that our simulations explore supercooled states close to the dynamical transition.
The shear rate $\dot{\gamma}$ is controlled over a wide range to cover the Newtonian to the strongly nonlinear regimes.
From the MD trajectory data, we measure the relaxation time $\tau_\alpha$, the diffusion constant $D$, and the viscosity $\eta$ as a function of $T$ and $\dot{\gamma}$.
For detailed information on MD simulations and calculations of $\tau_\alpha$, $D$, and $\eta$, please see Methods.

In addition to the GCM, we conduct MD simulations on the KA model under shear flow and measure $\tau_\alpha$ and $\eta$.
We also extract available data on the SS model from Ref.~\cite{Furukawa2_2023}, the two-dimensional SS~(2DSS) model from Ref.~\cite{Furukawa_2017}, and the BKS model from Ref.~\cite{Furukawa_2023}.
Details of MD simulations on the KA model and system descriptions of the SS, 2DSS, and BKS models are provided in Methods.

Figure~\ref{fig_data} presents the data on all the studied systems together.
We provide $\tau_\alpha$ for the GCM and the KA model and $\eta$ for the KA, SS, 2DSS, and BKS models.
Note that the linear relation $\tau_\alpha \propto \eta$ normally holds, as we confirm for the KA model in Methods and for the SS model in Ref.~\cite{Yamamoto2_1998}.
Thus, $\tau_\alpha$ and $\eta$ provide essentially the same information on dynamics, in general.
However, we find that $\tau_\alpha \propto \eta$ breaks at high shear rates in the GCM, as shown in Methods.
This point requires further detailed investigation.
In Fig.~\ref{fig_data}, we show $\tau_\alpha$ (not $\eta$) for the GCM.

In panel (a) of Fig.~\ref{fig_data}, we plot $\tau_\alpha/\tau_{\alpha 0}$ and $\eta/\eta_0$ against $\dot{\gamma}/\dot{\gamma}_c$.
Here, $\tau_\alpha$, $\eta$, and $\dot{\gamma}$ are normalized using the equilibrium values $\tau_{\alpha 0}$ and $\eta_0$, and the onset shear rate $\dot{\gamma}_c$, respectively, in order to compare different systems.
We observe that all the systems studied exhibit similar dependences on $\dot{\gamma}$, which are not proportional to $\dot{\gamma}^{-1}$~(red line), but rather proportional to $\dot{\gamma}^{-\nu}$ with $\nu \sim 0.7$~(blue line).
We particularly emphasize that the GCM does not follow the $\propto \dot{\gamma}^{-1}$ dependence of the advection scenario.
For the GCM, we also plot data on $D^{-1}/D_0^{-1}$ which are indistinguishable from those on $\tau_\alpha/\tau_{\alpha 0}$.
This agreement demonstrates that the SE law in the form of $\tau_{\alpha} \propto D^{-1}$ holds throughout the shear thinning regime, not just in equilibrium states~\cite{Ikeda_2011,Ikeda2_2011}.
Thus, both the structural relaxation and the diffusion dynamics of the GCM do not follow the $\propto \dot{\gamma}^{-1}$ of the advection scenario.

In panel (b) of Fig.~\ref{fig_data}, we present the onset shear rate $\dot{\gamma}_c$ as a function of $\tau_{\alpha 0}$ or $\eta_0$.
For all the systems studied, we observe that $\dot{\gamma}_c$ is much smaller than $\tau_{\alpha 0}^{-1}$ or $\eta_0^{-1}$~(red line), and it follows $\dot{\gamma}_c \propto \tau_{\alpha 0}^{-\delta}$ or $\propto \eta_0^{-\delta}$ with $\delta \sim 1.4$~(blue line).
In particular, the GCM does not follow $\dot{\gamma}_c \sim \tau_{\alpha 0}^{-1}$ of the advection scenario.
We thus conclude that the current sheared MCT fails to explain the shear thinning in supercooled liquids.
In contrast, the GCM and the other systems share similar scaling behaviors of $\tau_\alpha \propto \dot{\gamma}^{-\nu}$ with $\nu \sim 0.7$ and $\dot{\gamma}_c \propto \tau_{\alpha 0}^{-\delta}~(\ll \tau_{\alpha 0}^{-1})$ with $\delta \sim 1.4$.
This result suggests a universal mechanism of the shear thinning in supercooled liquids.

\begin{table}[t]
\caption{\label{table}
Dynamical transition temperature $T_c$, mode coupling theory~(MCT) scaling exponent $\gamma$, and predicted thinning exponents $\nu$ and $\delta$.
We present data for the Gaussian core model~(GCM), the Kob-Andersen~(KA) model, the soft sphere~(SS) model, the two-dimensional SS~(2DSS) model, and the van Beest-Kramer-van Santen~(BKS) model.
The values of $T_c$ and $\gamma$ are obtained from published papers cited in References.
For 2DSS, we determine $T_c$ and $\gamma$ by fitting the MCT power-law scaling to the data on $\eta_0$ versus $T$ published in Ref.~\cite{Furukawa_2017}.
After obtaining $\gamma$, we calculate $\nu$ and $\delta$ using the formulas $\nu = \gamma/(\gamma + 1)$ and $\delta = (\gamma+1)/\gamma$, respectively.
}
\centering
\renewcommand{\arraystretch}{1.3}
\begin{tabular}{c||c|c|c|c|c}
\hline
& GCM & KA & SS &\ 2DSS\ & BKS \\
\hline
\hline
References & \cite{Ikeda_2011,Ikeda2_2011} & \cite{Kob_1995,Kob2_1995,Berthier_2012,Kim_2013} & \cite{Berthier_2012,Kim_2013} & $-$ & \cite{Horbach_1999,Horbach_2001}  \\
\cline{1-6}
$\rho$ & $2.0$ & $1.2$ & $0.8$ & $0.8$ & $1.632$  \\
\cline{1-6}
$T_c$ & $2.68\times 10^{-6}$ & $0.435$ & $0.267$ & $0.534$ & $0.4775$  \\
\cline{1-6}
$\gamma$ & $2.7$ & $2.4$ & $2.2$ & $2.9$ & $2.4$ \\
\cline{1-6}
$\nu=\gamma/(\gamma + 1)$ & $0.73$ & $0.71$ & $0.69$ & $0.74$ & $0.71$ \\
\cline{1-6}
$\delta= (\gamma+1)/\gamma$ & $1.37$ & $1.42$ & $1.45$ & $1.34$ & $1.42$ \\
\hline
\end{tabular}
\end{table}

\subsection*{Distortion scenario}
As the advection scenario of the sheared MCT fails to explain the shear thinning, we turn our attention to the distortion scenario proposed by Furukawa~\cite{Furukawa_2017,Furukawa2_2023,Furukawa_2023}.
All the systems studied in Fig.~\ref{fig_data}, including not only the GCM but also the KA, SS, 2DSS, and BKS models, follow the MCT power-law scaling, Eq.~(\ref{MCTpower1}), in the temperature regime above the critical temperature $T_c$.
We summarize the values of $T_c$ and $\gamma$ in Table~\ref{table}, which are sourced from published papers.
Note that $\gamma$ in the current systems lies between $2$ and $3$.
This fact motivates us to explain a universal mechanism behind shear thinning by incorporating the distortion effect into the MCT power-law scaling.

The shear flow distorts the particles' configuration in an anisotropic manner, which causes the effective density $\rho_\text{eff}$ to increase along the compression axis and decrease along the decompression axis.
Since the distortion occurs on the timescale of $\tau_\alpha$, the variation of $\rho_\text{eff}$ due to the shear is characterized by the strain $\dot{\gamma}\tau_{\alpha}$.
Thus, assuming $\dot{\gamma} \tau_{\alpha} \ll 1$~(we confirm that $\dot{\gamma} \tau_{\alpha}$ is at most $10^{-1}$ for the GCM and the KA model), $\rho_\text{eff}$ can be described as
\begin{equation}
\rho_\text{eff} \approx \rho + b_\rho \dot{\gamma} \tau_{\alpha},
\label{distortion0}
\end{equation}
where $\rho$ is the density in the unsheared~(equilibrium) state, and ${\displaystyle b_\rho =\left. {\partial \rho_\text{eff}}/{\partial (\dot{\gamma}\tau_{\alpha})}\right|_{\dot{\gamma} \tau_{\alpha}=0}}$ can be positive or negative, depending on the direction of compression or decompression.  

We next consider how the variation in $\rho_\text{eff}$ impacts the relaxation time $\tau_{\alpha}$.
In the case of the GCM, since the dynamics accelerate as the density increases~\cite{Ikeda_2011,Ikeda2_2011}, the direction along the compression axis, where $b_\rho$ is positive and $\rho_\text{eff}$ increases, contributes to the shear thinning.
Although the dynamics become slow in the other direction~(along the decompression axis), this does not prevent the shear thinning because an acceleration along the compression axis leads to a significant acceleration in overall dynamics.
On the other hand, for the KA, SS, 2DSS, and BKS models, the dynamics speed up with decreasing the density, and the direction of the negative $b_\rho$ along the decompression axis causes the shear thinning.
This behavior is contrary to that of the GCM.
However, in both cases, a minute but finite variation in $\rho_\text{eff}$ commonly plays a crucial role in the shear thinning process.

Focusing on the direction of the positive $b_\rho$ for the GCM or that of the negative $b_\rho$ for the KA, SS, 2DSS, and BKS models, we can proceed with the following formulations for $\tau_\alpha(T,\dot{\gamma})$.
Recall that $\tau_{\alpha 0}(T)$ of the unsheared~(equilibrium) system follows the MCT power-law scaling, Eq.~(\ref{MCTpower1}), close to $T_c$; $\tau_{\alpha 0} \propto [T-T_c(\rho)]^{-\gamma}$ where $T_c$ is a function of the density $\rho$.
We assume that this power-law scaling remains valid under shear by replacing $\rho$ in $T_c$ by $\rho_\text{eff}$, {\it i.e.},
\begin{equation}
\tau_{\alpha} \propto [T-T_c(\rho_\text{eff})]^{-\gamma}.
\label{MCTpower1eff}
\end{equation}
In addition, applying Eq.~(\ref{distortion0}) for $\rho_\text{eff}$, we can approximate $T_c(\rho_\text{eff})$ as
\begin{equation}
T_c(\rho_\text{eff}) \approx T_c(\rho) - b_T \dot{\gamma} \tau_{\alpha},
\label{linearapprox}
\end{equation}
where ${\displaystyle b_T = -\left. \partial T_c/{\partial (\dot{\gamma}\tau_{\alpha})}\right|_{\dot{\gamma} \tau_{\alpha}=0} = -b_\rho \left. {dT_c}/{d\rho_\text{eff}}\right|_{\rho_\text{eff}=\rho}}$ is a positive constant regardless of the system, since $T_c(\rho_\text{eff})$ is a decreasing function of $\rho_\text{eff}$ for the GCM~\cite{Ikeda_2011,Ikeda2_2011} whereas it is an increasing function of $\rho_\text{eff}$ for the KA, SS, 2DSS, and BKS models.
Finally, using Eq.~(\ref{linearapprox}) for $T_c(\rho_\text{eff})$ in Eq.~(\ref{MCTpower1eff}), we arrive at a self-consistent equation for $\tau_\alpha(T,\dot{\gamma})$;
\begin{equation}
\tau_\alpha \propto \left[ T-T_c(\rho) + b_T \dot{\gamma} \tau_{\alpha}\right]^{-\gamma}.
\label{distortion2}
\end{equation}

By using Eq.~(\ref{distortion2}), we can make predictions for the onset shear rate and the thinning scaling as below.
Since $b_T \dot{\gamma} \tau_{\alpha 0}$ becomes comparable to $T-T_c(\rho) \propto \tau_{\alpha 0}^{-1/\gamma}$ at the onset $\dot{\gamma}=\dot{\gamma}_{c}$, we obtain
\begin{equation}
\dot{\gamma}_{c} \propto \tau_{\alpha 0}^{-\delta}, \quad \delta = \frac{\gamma+1}{\gamma}.
\label{firstonset}
\end{equation}
In addition, once the shear thinning builds up, $b_T \dot{\gamma} \tau_{\alpha} \gg T-T_c(\rho)$ holds.
Thus, Eq.~(\ref{distortion2}) leads to $\tau_\alpha \propto (b_T \dot{\gamma} \tau_{\alpha})^{-\gamma}$, giving a thinning scaling of
\begin{equation}
\tau_\alpha \propto \dot{\gamma}^{-\nu}, \quad \nu = \frac{\gamma}{\gamma+1}.
\label{firstscale}
\end{equation}

Note that for strong glass formers like the BKS model, we need to consider the activation energy $E_\text{eff}$ instead of the density $\rho_\text{eff}$ in the above formulations~\cite{Furukawa_2023}.
However, by replacing $\rho_\text{eff}$ with $E_\text{eff}$, we arrive at the same self-consistent equation for $\tau_\alpha(T,\dot{\gamma})$, Eq~(\ref{distortion2}).
This results in obtaining the same formulations as in Eqs.~(\ref{firstonset}) and~(\ref{firstscale}) for the strong glass formers.

We thus derive expressions for the thinning exponents $\nu = \gamma/(\gamma + 1)$ in Eq.~(\ref{firstscale}) and $\delta = (\gamma + 1)/\gamma$ in Eq.~(\ref{firstonset}).
These expressions are applicable to any system that remains above the dynamical transition temperature $T_c$ and follows the MCT power-law scaling given by Eq. (\ref{MCTpower1}).
By substituting specific values of $\gamma$ into these expressions, we can obtain values for $\nu$ and $\delta$, which are summarized in Table \ref{table}.
For the present systems, we have values of $\gamma$ ranging from $2$ to $3$, resulting in $\nu \sim 0.7$ and $\delta \sim 1.4$, which are quantitatively consistent with the observations in Fig.~\ref{fig_data}.

Therefore, we conclude that the distortion scenario universally works in different types of supercooled liquids, including the GCM and the fragile and strong glass formers.
The exponents, $\nu= \gamma/(\gamma+1)$ and $\delta=(\gamma+1)/\gamma$, are determined by the MCT exponent $\gamma$.
This means that the power-law scaling in the shear thinning comes from the equilibrium MCT critical scaling near $T_c$.
The present systems show similar thinning exponents $\nu~(\sim 0.7)$ and $\delta~(\sim 1.4)$, which are generated by similar values of $\gamma~(\sim 2$ to $3)$.

\begin{figure}[t]
\centering
\includegraphics[width=0.45\textwidth]{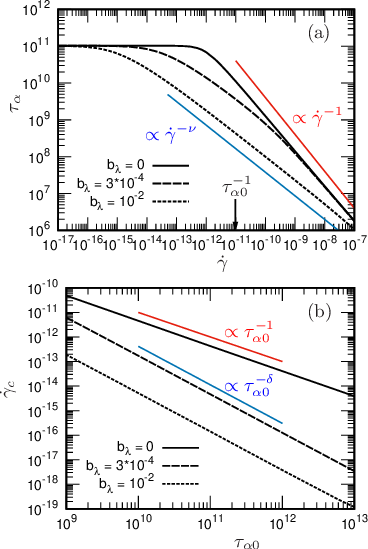}
\caption{\label{fig_mct}
Predictions of the extended mode coupling theory~(MCT) model.
We present predictions for different $b_\lambda$ values; $b_\lambda = 0$~(solid lines), $3 \times 10^{-4}$~(dashed lines), and $10^{-2}$~(dotted lines).
Note that $b_\lambda = 0$ corresponds to the original MCT model with advection only. 
(a) $\tau_\alpha$ is plotted as a function of $\dot{\gamma}$.
$\lambda$ is fixed at $\lambda_c - \lambda = 2\times 10^{-6}$~($\lambda_c=4$).
The arrow indicates $\dot{\gamma} = \tau_{\alpha 0}^{-1}$.
The red and blue lines respectively indicate $\tau_\alpha \propto \dot{\gamma}^{-1}$ due to advection and $\tau_\alpha \propto \dot{\gamma}^{-\nu}$ with $\nu = \gamma/(\gamma + 1) \simeq 0.64~(<1)$ due to distortion.
(b) $\dot{\gamma}_{c}$ is plotted against $\tau_{\alpha 0}$.
The red and blue lines respectively indicate $\dot{\gamma}_{c} \propto \tau_{\alpha 0}^{-1}$ due to advection and $\dot{\gamma}_{c} \propto \tau_{\alpha 0}^{-\delta}$ with $\delta = (\gamma + 1)/\gamma \simeq 1.57~(> 1)$ due to distortion.
}
\end{figure}

\subsection*{Schematic model of the sheared MCT}
As we have seen so far, the distortion scenario accompanied by the MCT power-law scaling is successful in explaining the nontrivial values of exponents $\nu \sim 0.7$ and $\delta \sim 1.4$ in observations, through $\nu = \gamma/(\gamma + 1)$ and $\delta = (\gamma +1)/\gamma$.
In the next step, we will integrate the distortion mechanism into the current sheared MCT to renovate the theory.

For this goal, we shall consider the schematic version of the MCT which drops wavenumber~($k$) dependences~\cite{MCT}. 
The schematic MCT for unsheared~(equilibrium) liquids, which is also known as the Leutheusser equation~\cite{Leutheusser_1984}, has the mathematically same form as the $k$-dependent full MCT and preserves key characteristics of nontrivial slow dynamics and the dynamical transition, such as the power-law divergence of the relaxation time~\cite{Gotze_1984,Franosch_1997}.
The schematic MCT has been extended to sheared liquids~\cite{Fuchs_2003,Brader_2009,Voigtmann_2012}, which again retains consequences of the $k$-dependent full sheared MCT.

We start with the sheared $F_2^{(\dot{\gamma})}$ model proposed by Fuchs and Cates~\cite{Fuchs_2003},
\begin{equation}
\dot{\phi}(t) + {\phi}(t) + \int_0^t m(t-s) \dot{\phi}(s) ds = 0,
\end{equation}
with the memory kernel,
\begin{equation}
m(t) = \frac{1}{1+(\dot{\gamma}t)^2} \lambda \phi^2(t),
\label{model0}
\end{equation}
where $\phi(t)$ represents a normalized intermediate scattering function, and the dot denotes the time derivative.
$\lambda$ is a parameter that contains information on the static structure factor and the temperature.
In the equilibrium states with $\dot{\gamma}=0$, this model predicts the MCT power-law scaling,
\begin{equation}
\tau_{\alpha 0}(\lambda) \propto (\lambda_c - \lambda)^{-\gamma},
\label{MCTpowerschematic}
\end{equation}
with $\gamma \simeq 1.76$, and the dynamical transition at $\lambda_c=4$.
The term $(\dot{\gamma}t)^2$ in the denominator in $m(t)$ of Eq.~(\ref{model0}) accounts for the advection effect by the shear flow. 
The fact that $\dot{\gamma}$ is scaled by $t$ demonstrates that the advection and its resultant decoupling of the nonlinear coupling of density fields are responsible for shear thinning.
As does the $k$-dependent full MCT, the model predicts $\tau_\alpha \propto \dot{\gamma}^{-1}$ and $\dot{\gamma}_{c} \propto \tau_{\alpha 0}^{-1}$, {\it i.e.}, values of exponents $\nu=1$ and $\delta =1$~(see solid lines of $b_\lambda = 0$ in Fig.~\ref{fig_mct}).
These predictions contradict the numerical observations of $\nu < 1$ and $\delta > 1$ in Fig.~\ref{fig_data}, which are, however, correctly captured by the distortion scenario~\cite{Furukawa_2017,Furukawa2_2023,Furukawa_2023}, as we have demonstrated in the previous section.

In the equilibrium MCT, the static structure factor $S(|\mathbf{k}|)$ (or $\lambda$ in the schematic version) is an essential input parameter~($\mathbf{k}$ is wavevector, and $|\mathbf{k}|=k$).
In the sheared systems, $S(|\mathbf{k}|)$ (or $\lambda$) is distorted and replaced by a nonequilibrium function $S_\text{NE}(\mathbf{k})$~\cite{Ronis_1984,Hanley_1987,Iwashita_2012,Yamaguchi_2018} (or $\lambda_{\text{NE}}$).
So far, the sheared MCT has never taken $S_\text{NE}(\mathbf{k})$ into account based on the observation that the distortion of $S(|\mathbf{k}|)$ to $S_\text{NE}(\mathbf{k})$ is very small~\cite{Miyazaki_2004,Fuchs_2009}.
However, we now understand from the distortion scenario that this tiny distortion is surely responsible for shear thinning and needs to be addressed in the theory.

Here, we propose to introduce the distortion effect into the schematic MCT model, Eq.~(\ref{model0}), by modifying $\lambda$ to $\lambda_{\text{NE}}$ as follows.
The procedure is the same in formulating Eq.~(\ref{distortion0}).
The distortion occurs on the timescale of the structural relaxation time $\tau_\alpha$, and the density field experiences the strain $\dot{\gamma}\tau_{\alpha}$.
Thus, assuming $\dot{\gamma}\tau_{\alpha} \ll 1$, the distorted parameter $\lambda_{\text{NE}}$ can be expressed as
\begin{equation}
\lambda_\text{NE} \approx \lambda- b_\lambda \dot{\gamma} \tau_{\alpha},
\label{distortion0schematic}
\end{equation}
where ${\displaystyle b_\lambda=\left. -{\partial \lambda_\text{NE}}/{\partial (\dot{\gamma}\tau_{\alpha})}\right|_{\dot{\gamma} \tau_{\alpha}=0}}~(>0)$ is a model parameter which quantifies sensitivity to the shear flow.
Replacing $\lambda$ by $\lambda_\text{NE}$ in Eq.~(\ref{model0}) while keeping the advection effect, we have
\begin{equation}
m(t) = \frac{1}{1+(\dot{\gamma}t)^2} (\lambda - b_\lambda \dot{\gamma} \tau_{\alpha})  \phi^2(t), 
\label{model2}
\end{equation}
which accounts for the distortion effect in addition to the advection effect.
In the case of $b_\lambda = 0$, the model reduces to the original model, Eq.~(\ref{model0}), with the advection effect only.
By setting $b_\lambda > 0$, the distortion effect is introduced, and it is increased by increasing $b_\lambda$.

Let us first discuss the distortion effect solely by considering the MCT model without the advection;
\begin{equation}
m(t) = (\lambda - b_\lambda \dot{\gamma} \tau_{\alpha})  \phi^2(t).
\label{model1}
\end{equation}
Since $\lambda$ is replaced by $\lambda_\text{NE} = \lambda - b_\lambda \dot{\gamma} \tau_{\alpha}$~[Eq.~(\ref{distortion0schematic})], the MCT power-law scaling, Eq.~(\ref{MCTpowerschematic}), for $\tau_{\alpha 0}(\lambda)$~(in the absence of $\dot{\gamma}$) transforms to a self-consistent equation for $\tau_\alpha(\lambda,\dot{\gamma})$;
\begin{equation}
\tau_{\alpha} \propto (\lambda_c - \lambda_\text{NE})^{-\gamma} = (\lambda_c - \lambda + b_\lambda \dot{\gamma} \tau_{\alpha})^{-\gamma}.
\label{model1a}
\end{equation}
This equation is essentially the same as Eq.~(\ref{distortion2}).
Solving Eq.~(\ref{model1a}), we obtain the same formulations of $\delta = (\gamma+1)/\gamma$ in Eq.~(\ref{firstonset}) and $\nu = \gamma/(\gamma+1)$ in Eq.~(\ref{firstscale}).
We thus conclude that the distortion effect is correctly embedded in the MCT model, Eq.~(\ref{model2}).
The MCT exponent $\gamma \simeq 1.76$ provides specific values of $\delta = (\gamma+1)/\gamma \simeq 1.57~(> 1)$ and $\nu = \gamma/(\gamma+1) \simeq 0.64~(< 1)$ in this MCT model.

Figure~\ref{fig_mct} presents numerical solutions of the MCT model, Eq.~(\ref{model2}), for three different values of $b_\lambda$: $0$~(solid lines), $3\times 10^{-4}$~(dashed lines), and $10^{-2}$~(dotted lines). 
Panel (a) shows $\tau_\alpha$ as a function of $\dot{\gamma}$ for $\lambda_c - \lambda = 2\times 10^{-6}$, while panel (b) shows $\dot{\gamma}_c$ as a function of $\tau_{\alpha 0}$.
When $b_\lambda = 0$, the advection scenario is produced, with $\tau_\alpha \propto \dot{\gamma}^{-1}$ and $\dot{\gamma}_c \sim \tau_{\alpha 0}^{-1}$. 
However, when $b_\lambda = 10^{-2}$, the distortion scenario is produced, with $\tau_\alpha \propto \dot{\gamma}^{-\nu}$ and $\dot{\gamma}_c \sim \tau_{\alpha 0}^{-\delta}$ as described in Eqs.~(\ref{firstscale}) and~(\ref{firstonset}), respectively. 
At the intermediate value of $b_\lambda = 3\times 10^{-4}$, the model predicts the crossover from the distortion-induced $\tau_\alpha \propto \dot{\gamma}^{-\nu}$ to the advection-induced $\tau_\alpha \propto \dot{\gamma}^{-1}$.
The model also explains that a larger distortion effect with increasing $b_\lambda$ results in a much slower $\dot{\gamma}_c$.

In the present systems of the GCM and the KA, SS, 2DSS, BKS models, the situations all relate to large values of $b_\lambda$.
In such cases, the primary factor affecting shear thinning is distortion rather than advection.
As a result, we update the sheared MCT to include the distortion effect, which resolves long-standing inconsistencies between the theory and the observations in experiments and simulations.

\section*{Conclusions}
We have addressed questions (i) to (iii) raised in the Introduction.
Firstly, we have observed that different types of systems, namely the GCM, and the KA, SS, and BKS models, exhibit similar scalings of $\tau_\alpha,\ \eta \propto \dot{\gamma}^{-\nu}$ with $\nu \sim 0.7~(< 1)$ and $\dot{\gamma}_c \propto \tau_{\alpha 0}^{-\delta}$ with $\delta \sim 1.4~(>1)$, as in Fig.~\ref{fig_data}.
The GCM does not follow $\tau_\alpha \propto \dot{\gamma}^{-1}$ and $\dot{\gamma}_c \propto \tau_{\alpha 0}^{-1}$ of the advection scenario, which dictates that (i) the current sheared MCT fails to explain the shear thinning.
Next, we used the distortion scenario accompanied by the MCT power-law scaling and formulated the thinning exponents, $\nu = \gamma/(\gamma + 1)$ in Eq.~(\ref{firstscale}) and $\delta = (\gamma + 1)/\gamma$ in Eq.~(\ref{firstonset}), in terms of the MCT exponent $\gamma$.
These formulations provide quantitatively correct values of $\nu \sim 0.7$ and $\delta \sim 1.4$ in the observations, thus concluding that (ii) the distortion scenario works universally in the GCM and the fragile and strong glass formers.
Finally, we integrated the distortion effect into the schematic MCT model as in Eq.~(\ref{model2}), which explains $\nu = \gamma/(\gamma + 1)$ and $\delta = (\gamma + 1)/\gamma$.
Consequently, (iii) we renovated the sheared MCT by accounting for the distortion effect.
Our numerical and theoretical results (i) to (iii) have resolved the long-standing discrepancies between the theory and the observations in experiments and simulations, establishing a universal mechanism of shear thinning in supercooled liquids.

The thinning exponents $\nu$ and $\delta$ are determined by the MCT exponent $\gamma$.
This indicates that the power-law scalings observed in shear thinning originate from the criticality of the equilibrium MCT near the dynamical transition point $T_c$.
All the systems studied in this work exhibit similar shear-rate dependences for $\tau_\alpha$ or $\eta$, which is due to their similar values of $\gamma$, ranging from $2$ to $3$.
It would be interesting in future research to investigate systems with values of $\gamma$ that differ significantly from this range.
For instance, the harmonic spheres can display $\gamma \simeq 5.3$ at high packing fractions above $\varphi = 0.8$~\cite{Berthier_2009}, resulting in $\nu \simeq 0.84$ and $\delta \simeq 1.19$.

On the other hand, although macroscopic observables~($\tau_\alpha$ and $\eta$) follow similar shear-rate dependences across different systems, microscopic dynamics are expected to be quite different.
The equilibrium dynamics of the GCM differ significantly from those of typical liquids with short-ranged, harshly-repulsive potentials like the KA model~\cite{Coslovich_2016}.
In the KA model, dynamics are described by the caging mechanism with hopping motions between local cages, whereas the GCM exhibits rather continuous motions that are not characterized by the standard caging mechanism.
The most recent work~\cite{Sposini_2023} reported that the GCM also exhibits the caging dynamics at low densities, and upon increasing the density, a smooth variation occurs towards the non-caging dynamics.
In addition, it was reported that dynamics are very different between fragile glass formers~(SS model) and strong glass formers~(BKS model)~\cite{Furukawa_2016}.
Therefore, one would expect that microscopic dynamics under shear flow differ significantly between the GCM and the fragile and strong glass formers.

The present work focuses on the temperature regime above the dynamical transition temperature $T_c$.
In this regime, the shear thinning is closely related to the equilibrium MCT criticality.
On the other hand, we expect a distinct behavior below $T_c$.
At the mean-field level, the equilibrium dynamics transition from non-activation to activation as the temperature decreases across $T_c$.
In finite dimensions, non-mean-field effects disrupt this transition, but we can still observe its remnants as a dynamical crossover in the KA model~\cite{Coslovich_2018,Das_2022}.
In the future, it would be interesting to explore the nonlinear rheology below the dynamical transition.

Finally, it is commonly accepted that the viscosity is proportional to the relaxation time as $\eta \propto \tau_\alpha$~(as shown in Methods for the KA model); the relaxation time is responsible for controlling the viscosity in glass-forming liquids. 
However, as shown in Methods, we have observed that this relationship does not apply in the GCM at high shear rates.
This suggests that the shear modulus, measured as $G = \eta/\tau_\alpha$, is dependent on the shear rate $\dot{\gamma}$; in the GCM, as $\dot{\gamma}$ increases, so does $G$.
Further analysis is required to investigate this matter in the future.

\section*{Methods}
%
\subsection*{MD simulations on GCM subjected to shear flow}
We perform MD simulations on the mono-disperse GCM in three spatial dimensions~\cite{Ikeda_2011,Ikeda1_2011,Ikeda2_2011,Coslovich_2016}.
The particles interact via the potential,
\begin{equation}
v(r) = \epsilon e^{-(r/\sigma)^2},
\end{equation}
where $\epsilon$ and $\sigma$ characterize energy and length scales, respectively.
The interaction is truncated at $r= 5 \sigma$.
The mass of particles is $m$.
We use $\sigma$, $\epsilon/k_B$~($k_B$ is Boltzmann constant), and $\tau = (m \sigma^2/\epsilon)^{1/2}$ as units of length, temperature, and time, respectively.
The number density is fixed at $\rho =N/V = 2.0$, where $N=4000$ is the number of particles and $V$ is the system volume.
At $\rho = 2.0$, the dynamical transition temperature estimated by the standard power-law fitting for $\tau_{\alpha0}$ is $T_c \simeq 2.68\times 10^{-6}$~\cite{Ikeda_2011,Ikeda2_2011}.
To explore supercooled states, we study various temperatures ranging from $T \times 10^6 = 7.0$ to $2.9$ which is close enough to $T_c$.

After the system was equilibrated at each temperature $T$, we applied a steady shear flow to drive the system into a nonequilibrium state~\cite{Yamamoto_1998,Berthier_2002,Furukawa_2009,Mizuno_2012}.
We integrated the SLLOD equations using the Lees-Edwards boundary condition, with the Nos$\acute{\text{e}}$-Hoover thermostat to maintain the temperature~\cite{Nonequ}.
To cover the Newtonian to the strongly nonlinear regimes, we control the shear rate $\dot{\gamma}$ over a wide range from $\dot{\gamma} =10^{-8}$ to $10^{-3}$.
Here we set the $x$ axis along the flow direction and the $y$ axis along the velocity gradient direction.
The mean velocity profile $\mathbf{v}$ is thus given as 
\begin{equation}
\mathbf{v}=\dot{\gamma} y \mathbf{e}_x,
\end{equation}
where $\mathbf{e}_\mu$~($\mu=x,y,z$) is the unit vector along the $\mu$ axis.
We note that $\dot{\gamma} \sim 5\times 10^{-4}$ is high enough that the relaxation time $\tau_\alpha$ reaches the timescale of vibrations, the so-called Einstein period~\cite{Mizuno_2013}.

\begin{figure}[t]
\centering
\includegraphics[width=0.425\textwidth]{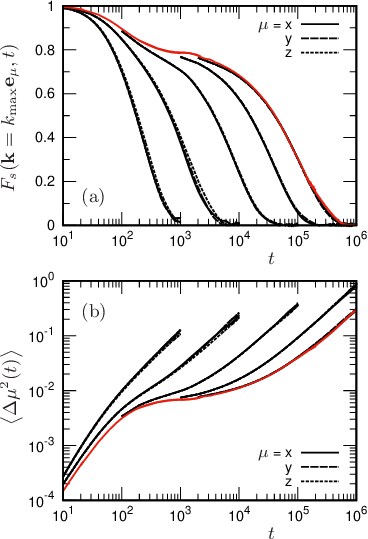}
\caption{\label{fig_fskt}
{Self-intermediate scattering function and mean squared displacements in the Gaussian core model~(GCM).}
(a) $F_s(\mathbf{k} = k_\text{max}\mathbf{e}_\mu,t)$ and (b) $\left< \Delta \mu^2(t) \right>$ are plotted as a function of $t$.
The temperature is $T\times 10^6 = 3.0$.
The red lines present values in the equilibrium state with $\dot{\gamma}=0$, while black lines give values in the sheared states with different $\dot{\gamma}$; from right to left, $\dot{\gamma}=10^{-7}$, $10^{-6}$, $10^{-5}$, $10^{-4}$, and $10^{-3}$.
We plot data for $\mu = x$, $y$, and $z$ directions or components, represented by solid, dashed, and dotted lines, respectively.
}
\end{figure}

\subsection*{Self-intermediate scattering function and mean squared displacements of GCM}
We employ two measurements to study the dynamics of particles; the self-intermediate scattering function $F_s(\mathbf{k},t)$,
\begin{equation}
F_s(\mathbf{k},t) = \left< \frac{1}{N} \sum_{i=1}^N e^{\text{i}\mathbf{k} \cdot \left[ \mathbf{r}_i(t) - \mathbf{r}_i(0) - \dot{\gamma}\int_0^t y_i(s) ds \mathbf{e}_x \right]} \right>,
\end{equation}
and the mean squared displacements $\left< \Delta \mathbf{r}^2(t) \right>$,
\begin{equation}
\left< \Delta \mathbf{r}^2(t) \right> = \left< \frac{1}{N} \sum_{i=1}^N \left[\mathbf{r}_i(t) - \mathbf{r}_i(0) - \dot{\gamma} \int_0^t y_i(s) ds \mathbf{e}_x \right]^2 \right>,
\end{equation}
where $\mathbf{r}_i=(x_i,y_i,z_i)$ is the position of particle $i$, $\left< \right>$ denotes the ensemble average, and we subtract from the total displacement of each particle, the contribution resulting from the advective transport by the mean shear flow, ${\displaystyle \dot{\gamma} \int_0^t y_i(s) ds \mathbf{e}_x}$~\cite{Yamamoto_1998}.

Figure~\ref{fig_fskt}(a) displays $F_s(\mathbf{k},t)$ for $\mathbf{k} = k_\text{max}\mathbf{e}_x$, $k_\text{max}\mathbf{e}_y$, and $k_\text{max}\mathbf{e}_z$, where $k_\text{max} \simeq 8.4$ is the wavenumber at which the static structure factor takes a maximum.
In Fig.~\ref{fig_fskt}(b), we report $\left< \Delta \mathbf{r}^2(t) \right>$ by separating $x$, $y$, and $z$ components;
\begin{equation}
\begin{aligned}
\left< \Delta {x}^2(t) \right> &= \left< \frac{1}{N} \sum_{i=1}^N \left[ {x}_i(t) - {x}_i(0) - \dot{\gamma} \int_0^t y_i(s) ds \right]^2 \right>, \\
\left< \Delta {y}^2(t) \right> &= \left< \frac{1}{N} \sum_{i=1}^N \left[ {y}_i(t) - {y}_i(0) \right]^2 \right>, \\
\left< \Delta {z}^2(t) \right> &= \left< \frac{1}{N} \sum_{i=1}^N \left[ {z}_i(t) - {z}_i(0) \right]^2 \right>.
\end{aligned}
\end{equation}
We observe that both $F_s(\mathbf{k},t)$ and $\left< \Delta \mathbf{r}^2(t) \right>$ show a drastic acceleration of the dynamics due to the shear flow.
In addition, both data are isotropic, showing little dependence on $x$, $y$, and $z$ directions or components, even at the highest shear rate $\dot{\gamma} = 10^{-3}$.

\begin{figure}[t]
\centering
\includegraphics[width=0.485\textwidth]{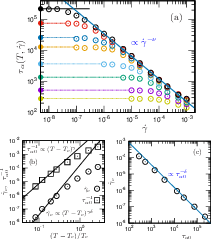}
\caption{\label{fig_gcm}
{Shear-rate dependence of the relaxation time and the onset shear rate of the shear thinning in the Gaussian core model~(GCM).}
(a) $\tau_\alpha(T,\dot{\gamma})$ is plotted as a function of $\dot{\gamma}$.
Symbols of different colors represent values at different $T$; from bottom to top, $T \times 10^{6} = 10.0$~(yellow), $7.0$~(purple), $5.0$~(green), $4.0$~(cyan), $3.4$~(orange), $3.2$~(blue), $3.0$~(red), and $2.9$~(black), all of which are above $T_c \times 10^{6} \simeq 2.68$.
Closed symbols indicate the equilibrium values $\tau_{\alpha 0}(T)$.
For $T\times 10^{6}=2.9$, black line indicates $\tau_\alpha = \tau_{\alpha 0} \propto \dot{\gamma}^0$, whereas blue line indicates $\tau_\alpha \propto \dot{\gamma}^{-\nu}$ with $\nu \simeq 0.73$~[Eq.~(\ref{firstscale})].
(b) $\dot{\gamma}_{c}$ and $\tau_{\alpha 0}^{-1}$ are plotted as a function of $(T-T_c)/T_c$.
The lines present the power-law scalings, $\tau_{\alpha 0}^{-1} \propto (T-T_c)^{\gamma}$ with $\gamma \simeq 2.7$~[Eq.~(\ref{MCTpower1})] and $\dot{\gamma}_{c} \propto (T-T_c)^{\gamma \delta}$ with $\gamma \delta \simeq 3.7$~[Eq.~(\ref{MCTpower2})].
(c) $\dot{\gamma}_{c}$ is plotted against $\tau_{\alpha 0}$.
The blue line presents the scaling relation of $\dot{\gamma}_{c} \propto \tau_{\alpha 0}^{-\delta}$ with $\delta \simeq 1.37$~[Eq.~(\ref{firstonset})].
}
\end{figure}

\begin{figure}[t]
\centering
\includegraphics[width=0.425\textwidth]{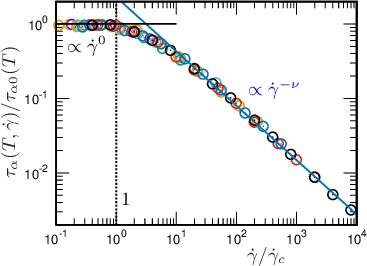}
\caption{\label{fig_gcm2}
{Scaled plot for shear-rate dependence of the relaxation time in the Gaussian core model~(GCM).}
$\tau_\alpha(T,\dot{\gamma})/\tau_{\alpha 0}(T)$ is plotted as a function of $\dot{\gamma}/\dot{\gamma}_{c}$.
Symbols of different colors represent values at different $T$; $T \times 10^{6} = 10.0$~(yellow), $7.0$~(purple), $5.0$~(green), $4.0$~(cyan), $3.4$~(orange), $3.2$~(blue), $3.0$~(red), and $2.9$~(black).
Black and blue lines indicate $\tau_\alpha/\tau_{\alpha 0} = 1$ and $\tau_\alpha/\tau_{\alpha 0} \propto (\dot{\gamma}/\dot{\gamma}_{c})^{-\nu}$ with $\nu \simeq 0.73$, respectively.
The vertical dotted line indicates the location of the onset shear rate $\dot{\gamma}_{c}$.
}
\end{figure}

\subsection*{Relaxation time of GCM}
From the relaxation behavior of $F_s(\mathbf{k},t)$, we calculate the relaxation time $\tau_\alpha$ as
\begin{equation}
F_s(|\mathbf{k}|=k_\text{max},t=\tau_\alpha) = e^{-1}.
\end{equation}
Figure~\ref{fig_gcm}(a) shows the shear-rate $\dot{\gamma}$ dependence of $\tau_\alpha(T,\dot{\gamma})$ for various temperatures $T$.
The figure demonstrates the shear thinning behavior in the GCM, which is characterized by a power-law scaling $\tau_\alpha\propto \dot{\gamma}^{-\nu}$ with an exponent $\nu \simeq 0.73$~(blue line).
The value of $\nu \simeq 0.73$ is obtained through $\nu = \gamma/(\gamma + 1)$ with $\gamma = 2.7$ as shown in Table~\ref{table}.
We note that at the high shear rate $\dot{\gamma} \gtrsim 5\times 10^{-4}$, $\tau_\alpha$ reaches the timescale of vibrations~\cite{Mizuno_2013}, and consequently it deviates from the scaling behavior of $\tau_\alpha\propto \dot{\gamma}^{-\nu}$.

Next, for each temperature $T$, we measure the onset shear rate $\dot{\gamma}_{c}$ at which the shear thinning starts.
In Fig.~\ref{fig_gcm}(b), we plot $\dot{\gamma}_{c}$ as a function of $(T-T_c)/T_c$, and compare it to $\tau_{\alpha 0}^{-1}$ as the sheared MCT predicts $\dot{\gamma}_{c1} \sim \tau_{\alpha 0}^{-1}$.
It is observed that $\dot{\gamma}_{c}$ is considerably~(orders of magnitude) smaller than $\tau_{\alpha 0}^{-1}$.
Note that this figure also confirms the MCT power-law scaling, Eq.~(\ref{MCTpower1}), close to the critical temperature $T_c$~(see the line for squares), in keeping with previous works~\cite{Ikeda_2011,Ikeda2_2011}.
In addition, we display $\dot{\gamma}_{c}$ against $\tau_{\alpha 0}$ in Fig.~\ref{fig_gcm}(c), showing that the data are well fitted by $\dot{\gamma}_{c} \propto \tau_{\alpha 0}^{-\delta}$ with $\delta \simeq 1.37$~(blue line).
The value of $\delta \simeq 1.37$ is obtained through $\delta = (\gamma + 1)/\gamma$ with $\gamma = 2.7$ as in Table~\ref{table}.
This result suggests that $\dot{\gamma}_{c}$ follows a power-law scaling,
\begin{equation}
\dot{\gamma}_{c} \propto \tau_{\alpha 0}^{-\delta} \propto (T-T_c)^{\gamma \delta},
\label{MCTpower2}
\end{equation}
with $\gamma \delta \simeq 3.7$.
This scaling indeed works close to $T_c$, as confirmed in Fig.~\ref{fig_gcm}(b)~(see the line for circles).

We then present a scaled plot of $\tau_\alpha(T,\dot{\gamma})/\tau_{\alpha 0}(T)$ versus $\dot{\gamma}/\dot{\gamma}_{c}$ in Fig.~\ref{fig_gcm2}.
Note that in Fig.~\ref{fig_gcm2}, we exclude data at the high shear rates $\dot{\gamma} \ge 5\times 10^{-4}$ at which $\tau_\alpha$ reaches the timescale of vibrations.
The data collapse onto a single curve regardless of temperature, which establishes
\begin{equation} \label{scaling}
\frac{\tau_\alpha(T,\dot{\gamma})}{\tau_{\alpha 0}(T)}
\left\{ 
\begin{aligned}
& = 1 & (\dot{\gamma} \lesssim \dot{\gamma}_{c}), \\
& \propto \left( \frac{\dot{\gamma}}{\dot{\gamma}_{c}} \right)^{-\nu} & (\dot{\gamma} \gg \dot{\gamma}_{c}),
\end{aligned} 
\right.
\end{equation}
where $\nu \simeq 0.73$, and $\dot{\gamma}_{c} \propto \tau_{\alpha 0}^{-\delta}$ with $\delta \simeq 1.37$.
Figure~\ref{fig_data} in the main text presents data on $\tau_\alpha/\tau_{\alpha 0}$ versus $\dot{\gamma}/\dot{\gamma}_{c}$ at the lowest $T = 2.9 \times 10^{-6}$ in (a), and $\dot{\gamma}_c$ versus $\tau_{\alpha 0}$ in (b).

\begin{figure}[t]
\centering
\includegraphics[width=0.425\textwidth]{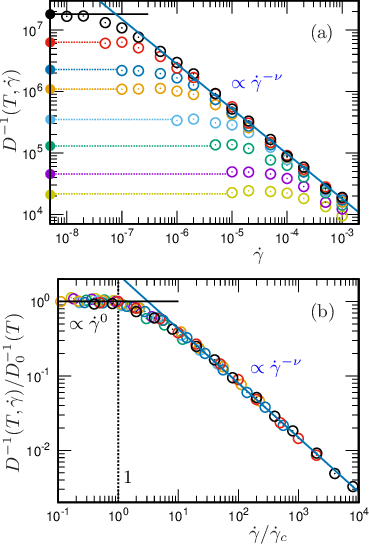}
\caption{\label{fig_gcmd}
{Shear-rate dependence of the (inverse) diffusion constant in the Gaussian core model~(GCM).}
(a) $D^{-1}(T,\dot{\gamma})$ is plotted as a function of $\dot{\gamma}$, and (b) $D^{-1}(T,\dot{\gamma})/D_0^{-1}(T)$ is plotted as a function of $\dot{\gamma}/\dot{\gamma}_{c}$.
Symbols of different colors represent values at different $T$; $T \times 10^{6} = 10.0$~(yellow), $7.0$~(purple), $5.0$~(green), $4.0$~(cyan), $3.4$~(orange), $3.2$~(blue), $3.0$~(red), and $2.9$~(black).
In (a), closed symbols indicate the equilibrium values $D_0^{-1}(T)$.
Black line indicates $D^{-1} = D^{-1}_0 \propto \dot{\gamma}^0$, whereas blue line indicates $D^{-1} \propto \dot{\gamma}^{-\nu}$ with $\nu \simeq 0.73$.
In (b), the vertical dotted line indicates the location of $\dot{\gamma}_{c}$.
}
\end{figure}

\subsection*{Diffusion constant of GCM}
The diffusion constant $D$ is determined by observing the diffusive behavior of $\left< \Delta \mathbf{r}^2(t) \right>$ in the long-time limit, which can be quantified as
\begin{equation}
\left< \Delta \mathbf{r}^2(t) \right> = 6D t.
\end{equation}
We present data on the inverse diffusion constant $D^{-1}(T,\dot{\gamma})$ in Fig.~\ref{fig_gcmd}(a) and $D^{-1}(T,\dot{\gamma})/D_0^{-1}(T)$ in Fig.~\ref{fig_gcmd}(b), which are counterparts of Figs.~\ref{fig_gcm}(a) and~\ref{fig_gcm2} for $\tau_\alpha(T,\dot{\gamma})$, respectively.
It is clear that $D^{-1}(T,\dot{\gamma})$ follows the same power-law behavior as that of $\tau_\alpha(T,\dot{\gamma})$ in Eq.~(\ref{scaling}).
In Fig.~\ref{fig_data} in the main text, we plot data on $D^{-1}/D^{-1}_{0}$ versus $\dot{\gamma}/\dot{\gamma}_{c}$ at $T = 2.9 \times 10^{-6}$, which are indistinguishable to those on $\tau_\alpha/\tau_{\alpha 0}$.
These observations demonstrate that the SE law in the form of $\tau_{\alpha} \propto D^{-1}$ holds throughout the shear thinning regime, not just in equilibrium states~\cite{Ikeda_2011,Ikeda2_2011}.

\begin{figure}[t]
\centering
\includegraphics[width=0.425\textwidth]{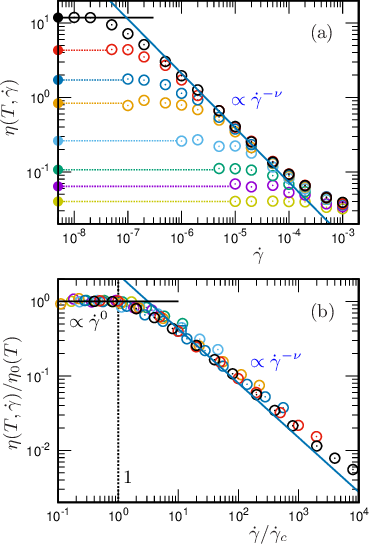}
\caption{\label{fig_gcmvisco}
{Shear-rate dependence of the viscosity in the Gaussian core model~(GCM).}
(a) $\eta(T,\dot{\gamma})$ is plotted as a function of $\dot{\gamma}$, and (b) $\eta(T,\dot{\gamma})/\eta_0(T)$ is plotted as a function of $\dot{\gamma}/\dot{\gamma}_{c}$.
Symbols of different colors represent values at different $T$; $T \times 10^{6} = 10.0$~(yellow), $7.0$~(purple), $5.0$~(green), $4.0$~(cyan), $3.4$~(orange), $3.2$~(blue), $3.0$~(red), and $2.9$~(black).
In (a), closed symbols indicate the equilibrium values $\eta_0(T)$.
Black line indicates $\eta = \eta_0 \propto \dot{\gamma}^0$, whereas blue line indicates $\eta \propto \dot{\gamma}^{-\nu}$ with $\nu \simeq 0.73$.
Note that the blue line, which is well fitted to data on $\tau_\alpha$ and $D^{-1}$, does not work for $\eta$.
In (b), the vertical dotted line indicates the location of $\dot{\gamma}_{c}$.
}
\end{figure}

\begin{figure}[t]
\centering
\includegraphics[width=0.425\textwidth]{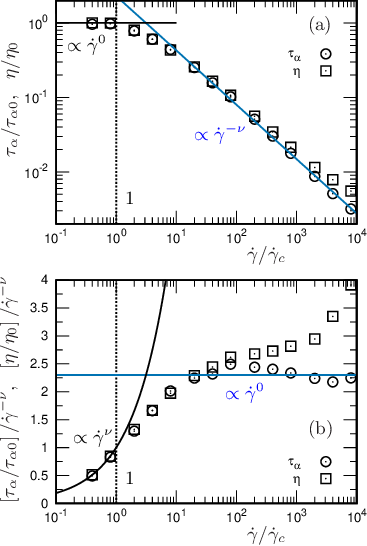}
\caption{\label{fig_gcmcom}
{Comparison of the relaxation time to the viscosity in the Gaussian core model~(GCM).}
(a) $\tau_\alpha(T,\dot{\gamma})/\tau_{\alpha 0}(T),\ \eta(T,\dot{\gamma})/\eta_{0}(T)$ and (b) $\left[\tau_\alpha(T,\dot{\gamma})/\tau_{\alpha 0}(T) \right]/\dot{\gamma}^{-\nu},\ \left[\eta(T,\dot{\gamma})/\eta_{0}(T) \right]/\dot{\gamma}^{-\nu}$ are plotted as a function of $\dot{\gamma}/\dot{\gamma}_{c}$.
Circles and squares represent values of $\tau_\alpha$ and $\eta$, respectively.
The temperature is $T\times 10^{6}=2.9$.
Black line indicates $\tau_\alpha/\tau_{\alpha 0},\ \eta/\eta_0 = 1$, whereas blue line indicates $\tau_\alpha/\tau_{\alpha 0},\ \eta/\eta_0 \propto \dot{\gamma}^{-\nu}$ with $\nu \simeq 0.73$.
The vertical dotted line indicates the location of $\dot{\gamma}_{c}$.
}
\end{figure}

\subsection*{Viscosity of GCM}
We measure the viscosity $\eta$ as a function of $T$ and $\dot{\gamma}$.
We calculate the shear stress $\sigma_{xy}$ as~\cite{Simpleliquid}
\begin{equation}
\sigma_{xy} = \left< -\frac{1}{V} \sum_{i=1}^N m v_{ix} v_{iy} + \frac{1}{V} \sum_{i=1}^{N-1} \sum_{j=i+1}^N \frac{d v(r_{ij})}{dr_{ij}} \frac{x_{ij} y_{ij}}{r_{ij}} \right>,
\end{equation}
where $\mathbf{v}_i=(v_{ix},v_{iy},v_{iz})$ is the velocity of particle $i$, $\mathbf{r}_{ij}=(x_{ij},y_{ij},z_{ij})$ denotes the vector $\mathbf{r}_i - \mathbf{r}_j=(x_i-x_j,y_i-y_j,z_i-z_j)$, and $r_{ij} = |\mathbf{r}_{ij}|$.
The viscosity is then obtained through $\eta = \sigma_{xy} / \dot{\gamma}$.

We present data on the viscosity $\eta(T,\dot{\gamma})$ in Fig.~\ref{fig_gcmvisco}(a) and $\eta(T,\dot{\gamma})/\eta_0(T)$ in Fig.~\ref{fig_gcmvisco}(b), which are counterparts of Figs.~\ref{fig_gcm}(a) and~\ref{fig_gcm2} for $\tau_\alpha(T,\dot{\gamma})$, respectively.
Note that the equilibrium values $\eta_0(T)$ are obtained by averaging values of $\eta(T,\dot{\gamma})$ in the Newtonian regime.
Although data on $\tau_\alpha/\tau_{\alpha 0}$ in Fig.~\ref{fig_gcm2}~(and $D^{-1}/D^{-1}_0$ in Fig.~\ref{fig_gcmd}(b)) are well fitted by $\propto (\dot{\gamma}/\dot{\gamma}_c)^{-\nu}$ with $\nu \simeq 0.73$ as mentioned in Eq.~(\ref{scaling}), it does not work for $\eta/\eta_0$ as we can observe in Fig.~\ref{fig_gcmvisco}(b).

In addition, Figure~\ref{fig_gcmcom} shows a comparison between $\eta/\eta_0$ and $\tau_\alpha/\tau_{\alpha 0}$ as a function of $\dot{\gamma}/\dot{\gamma}_c$.
At low shear rates of $\dot{\gamma}/\dot{\gamma}_c \lesssim 10^2$, we observe that $\eta/\eta_0$ coincides with $\tau_\alpha/\tau_{\alpha 0}$, which confirms that $\eta$ is proportional to $\tau_\alpha$.
However, this linear relation $\eta \propto \tau_\alpha$ is systematically violated at high shear rates of $\dot{\gamma}/\dot{\gamma}_c \gtrsim 10^2$.
These high shear rates correspond to the power-law scaling regime of $\tau_\alpha \propto \dot{\gamma}^{-\nu}$.

In many previous works~\cite{Yamamoto_1998,Berthier_2002,Furukawa_2009,Mizuno_2012,Furukawa_2017,Furukawa2_2023,Furukawa_2023}, it has been assumed that the relaxation time controls the viscosity in glass-forming liquids, and that $\eta \propto \tau_\alpha$.
This assumption has been confirmed for the KA model below and for the SS model in Ref.~\cite{Yamamoto2_1998}.
The sheared MCT also formulates $\eta \propto \tau_\alpha$~\cite{Fuchs_2002,Miyazaki_2002,Fuchs_2003,Miyazaki_2004}.
Thus, it is considered that $\tau_\alpha$ and $\eta$ provide essentially the same information on dynamics in supercooled liquids.
However, this assumption does not hold true for the GCM at high shear rates.
This result suggests that the shear modulus measured as $G = \eta/\tau_\alpha$ is dependent on $\dot{\gamma}$; $G$ increases as $\dot{\gamma}$ gets larger.
Further detailed investigation is required in the future to understand this point better.

\begin{figure}[t]
\centering
\includegraphics[width=0.485\textwidth]{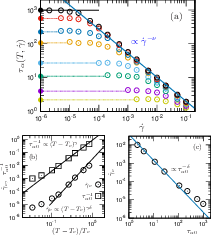}
\caption{\label{fig_kam}
{Shear-rate dependence of the relaxation time and the onset shear rate of the shear thinning in the Kob-Andersen~(KA) model.}
(a) $\tau_\alpha(T,\dot{\gamma})$ is plotted as a function of $\dot{\gamma}$.
Symbols of different colors represent values at different $T$; from bottom to top, $T = 0.8$~(yellow), $0.7$~(purple), $0.6$~(green), $0.55$~(cyan), $0.5$~(orange), $0.48$~(blue), $0.46$~(red), and $0.45$~(black), all of which are above $T_c \simeq 0.435$.
Closed symbols indicate the equilibrium values $\tau_{\alpha 0}(T)$.
For $T=0.45$, black line indicates $\tau_\alpha = \tau_{\alpha 0} \propto \dot{\gamma}^0$, whereas blue line indicates $\tau_\alpha \propto \dot{\gamma}^{-\nu}$ with $\nu \simeq 0.71$~[Eq.~(\ref{firstscale})].
(b) $\dot{\gamma}_{c}$ and $\tau_{\alpha 0}^{-1}$ are plotted as a function of $(T-T_c)/T_c$.
The lines present the power-law scalings, $\tau_{\alpha 0}^{-1} \propto (T-T_c)^{\gamma}$ with $\gamma \simeq 2.4$~[Eq.~(\ref{MCTpower1})] and $\dot{\gamma}_{c} \propto (T-T_c)^{\gamma \delta}$ with $\gamma \delta \simeq 3.4$~[Eq.~(\ref{MCTpower2})].
(c) $\dot{\gamma}_{c}$ is plotted against $\tau_{\alpha 0}$.
The blue line presents the scaling relation of $\dot{\gamma}_{c} \propto \tau_{\alpha 0}^{-\delta}$ with $\delta \simeq 1.42$~[Eq.~(\ref{firstonset})].
}
\end{figure}

\begin{figure}[t]
\centering
\includegraphics[width=0.425\textwidth]{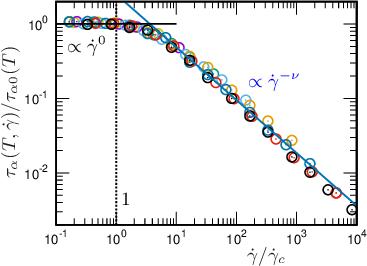}
\caption{\label{fig_kam2}
{Scaled plot for shear-rate dependence of the relaxation time in the Kob-Andersen~(KA) model.}
$\tau_\alpha(T,\dot{\gamma})/\tau_{\alpha 0}(T)$ is plotted as a function of $\dot{\gamma}/\dot{\gamma}_{c}$.
Symbols of different colors represent values at different $T$; $T = 0.8$~(yellow), $0.7$~(purple), $0.6$~(green), $0.55$~(cyan), $0.5$~(orange), $0.48$~(blue), $0.46$~(red), and $0.45$~(black).
Black and blue lines indicate $\tau_\alpha/\tau_{\alpha 0} = 1$ and $\tau_\alpha/\tau_{\alpha 0} \propto (\dot{\gamma}/\dot{\gamma}_{c})^{-\nu}$ with $\nu \simeq 0.71$, respectively.
The vertical dotted line indicates the location of the onset shear rate $\dot{\gamma}_{c}$.
}
\end{figure}

\begin{figure}[t]
\centering
\includegraphics[width=0.425\textwidth]{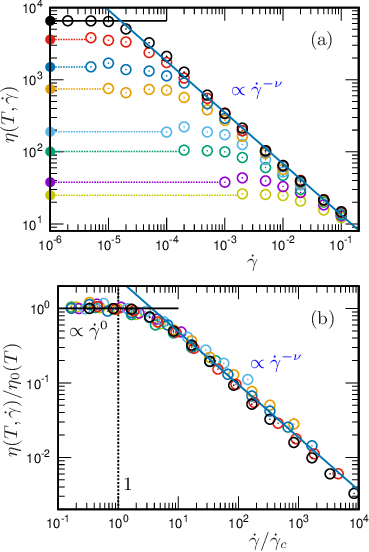}
\caption{\label{fig_kamvisco}
{Shear-rate dependence of the viscosity in the Kob-Andersen~(KA) model.}
(a) $\eta(T,\dot{\gamma})$ is plotted as a function of $\dot{\gamma}$, and (b) $\eta(T,\dot{\gamma})/\eta_0(T)$ is plotted as a function of $\dot{\gamma}/\dot{\gamma}_{c}$.
Symbols of different colors represent values at different $T$; $T = 0.8$~(yellow), $0.7$~(purple), $0.6$~(green), $0.55$~(cyan), $0.5$~(orange), $0.48$~(blue), $0.46$~(red), and $0.45$~(black).
In (a), closed symbols indicate the equilibrium values $\eta_0(T)$.
Black line indicates $\eta = \eta_0 \propto \dot{\gamma}^0$, whereas blue line indicates $\eta \propto \dot{\gamma}^{-\nu}$ with $\nu \simeq 0.71$.
In (b), the vertical dotted line indicates the location of $\dot{\gamma}_{c}$.
}
\end{figure}

\begin{figure}[t]
\centering
\includegraphics[width=0.425\textwidth]{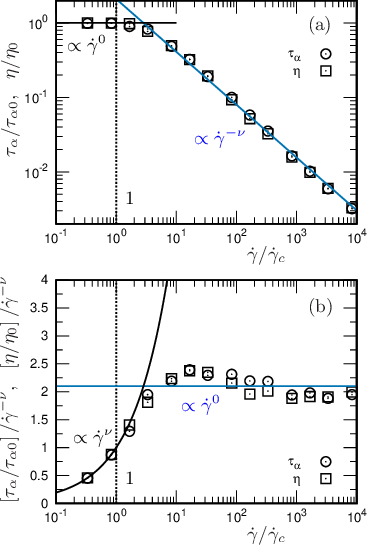}
\caption{\label{fig_kamcom}
{Comparison of the relaxation time to the viscosity in the Kob-Andersen~(KA) model.}
(a) $\tau_\alpha(T,\dot{\gamma})/\tau_{\alpha 0}(T),\ \eta(T,\dot{\gamma})/\eta_{0}(T)$ and (b) $\left[\tau_\alpha(T,\dot{\gamma})/\tau_{\alpha 0}(T) \right]/\dot{\gamma}^{-\nu},\ \left[\eta(T,\dot{\gamma})/\eta_{0}(T) \right]/\dot{\gamma}^{-\nu}$ are plotted as a function of $\dot{\gamma}/\dot{\gamma}_{c}$.
Circles and squares represent values of $\tau_\alpha$ and $\eta$, respectively.
The temperature is $T = 0.45$.
Black line indicates $\tau_\alpha/\tau_{\alpha 0},\ \eta/\eta_0 = 1$, wheares blue line indicates $\tau_\alpha/\tau_{\alpha 0},\ \eta/\eta_0 \propto (\dot{\gamma}/\dot{\gamma}_{c})^{-\nu}$ with $\nu \simeq 0.71$.
The vertical dotted line indicates the location of $\dot{\gamma}_{c}$.
}
\end{figure}

\subsection*{MD simulations on KA model subjected to shear flow}
We perform MD simulations on a binary Lennard-Jones~(LJ) mixture, the KA model, in three spatial dimensions~\cite{Kob_1994}.
The KA model is composed of large~($A$) and small~($B$) particles of equal masses, $m_A = m_B = m$.
The particles interact via the LJ potential,
\begin{equation}
v_{\alpha \beta}(r) = 4\epsilon_{\alpha \beta} \left[\left(\frac{\sigma_{\alpha \beta}}{r}\right)^{12} - \left(\frac{\sigma_{\alpha \beta}}{r}\right)^{6}\right],
\end{equation}
where $\alpha$ and $\beta$ denote $A$ or $B$, and the parameters are set to be $\epsilon_{AA} = \epsilon$, $\epsilon_{AB} = 1.5\epsilon_{AA}$, $\epsilon_{BB} = 0.5\epsilon_{AA}$, $\sigma_{AA} = \sigma$, $\sigma_{AB} = 0.8\sigma_{AA}$, $\sigma_{BB} = 0.88\sigma_{AA}$.
The interaction is truncated at $r= 2.5 \sigma_{\alpha \beta}$.
We employ $\sigma$, $\epsilon/k_B$, and $\tau = (m \sigma^2/\epsilon)^{1/2}$ as units of length, temperature, and time, respectively.
The number density is fixed at $\rho =N/V = 1.2$, and the number of particles is $N=N_A+N_B=4000$ with $N_A=3200$ and $N_B=800$.
At $\rho = 1.2$, the standard power-law fitting for $\tau_{\alpha0}$ estimates the dynamical transition temperature to be $T_c \simeq 0.435$~\cite{Kob_1994,Kob_1995,Kob2_1995}.
We study at various temperatures ranging from $T = 0.8$ to $0.45$ close to $T_c$.
The shear rate $\dot{\gamma}$ is controlled over a wide range of $\dot{\gamma} =10^{-6}$ to $10^{-1}$.
Note that $\dot{\gamma} \sim 10^{-1}$ is high enough that $\tau_\alpha$ reaches the timescale of vibrations~(Einstein period)~\cite{Mizuno_2013}.

We analyze the KA model in the same way as we do for the GCM.
At each temperature $T$, we measure the relaxation time $\tau_\alpha$ as a function of $\dot{\gamma}$ and identify the onset shear rate $\dot{\gamma}_c$.
We present results for the larger particles ($A$) below, but similar results were obtained for the smaller particles ($B$).
Figure~\ref{fig_kam} presents $\tau_{\alpha}$ versus $\dot{\gamma}$ in (a), and the data on $\dot{\gamma}_c$ and $\tau_{\alpha 0}^{-1}$ in (b) and (c).
Figure~\ref{fig_kam2} then plots $\tau_\alpha(T,\dot{\gamma})/\tau_{\alpha 0}(T)$ as a function of $\dot{\gamma}/\dot{\gamma}_{c}$, where we exclude data at $\dot{\gamma} = 10^{-1}$ at which $\tau_\alpha$ reaches the timescale of vibrations.
We also measure the viscosity $\eta$ and present data on $\eta(T,\dot{\gamma})$ and $\eta(T,\dot{\gamma})/\eta_0(T)$ in Fig.~\ref{fig_kamvisco}. Furthermore, we compare the relaxation time and the viscosity in Fig.~\ref{fig_kamcom}.
Figures~\ref{fig_kam}, \ref{fig_kam2}, \ref{fig_kamvisco}, and~\ref{fig_kamcom} for the KA model are counterparts of Figs.~\ref{fig_gcm}, \ref{fig_gcm2}, \ref{fig_gcmvisco}, and~\ref{fig_gcmcom} for the GCM, respectively.

In Figs.~\ref{fig_kam} and~\ref{fig_kam2}, we can see that the relaxation time follows a power-law scaling of the form $\tau_\alpha \propto \dot{\gamma}^{-\nu}$, where $\nu \simeq 0.71$, and the onset shear rate $\dot{\gamma}_{c}$ exhibits the scaling behavior $\dot{\gamma}_c \propto \tau_{\alpha 0}^{-\delta} \ll \tau_{\alpha 0}^{-1}$, with $\delta \simeq 1.42$.
The values of $\nu$ and $\delta$ are obtained using $\nu = \gamma/(\gamma + 1)$ and $\delta = (\gamma + 1)/\gamma$ with $\gamma = 2.4$ as shown in Table~\ref{table}.

In addition, Figures~\ref{fig_kamvisco} and~\ref{fig_kamcom} show that $\eta$ is proportional to $\tau_\alpha$ as $\eta \propto \tau_\alpha$, following the same scaling law as that of $\tau_\alpha$;
\begin{equation}
\frac{\eta(T,\dot{\gamma})}{\eta_0(T)}
\left\{ 
\begin{aligned}
&  =1 & (\dot{\gamma} \lesssim \dot{\gamma}_{c}), \\
&  \propto \left( \frac{\dot{\gamma}}{\dot{\gamma}_{c}} \right)^{-\nu} & (\dot{\gamma} \gg \dot{\gamma}_{c}).
\end{aligned} 
\right.
\end{equation}
Differently from the case of the GCM, the linear relation $\eta \propto \tau_\alpha$ is kept even at the high shear rates of $\dot{\gamma}/\dot{\gamma}_c \gtrsim 10^2$.
This result indicates that the relaxation time controls the viscosity in the KA model, as we normally expect and suppose.
Figure~\ref{fig_data} in the main text presents data on $\tau_\alpha/\tau_{\alpha 0}$ and $\eta/\eta_0$ versus $\dot{\gamma}/\dot{\gamma}_{c}$ at $T = 0.45$ in (a), and $\dot{\gamma}_c$ versus $\tau_{\alpha 0}$ in (b).

\subsection*{System description of SS model}
The SS model is a binary mixture composed of large~($L$) and small~($S$) particles in three spatial dimensions~\cite{Bernu_1985}.
The particles interact via the inverse power-law potential,
\begin{equation} \label{potss}
v_{\alpha \beta}(r) = \epsilon \left(\frac{\sigma_{\alpha \beta}}{r}\right)^{12},
\end{equation}
where $\alpha$ and $\beta$ denote $L$ or $S$, and $\sigma_{\alpha \beta} = (\sigma_\alpha + \sigma_\beta)/2$ with $\sigma_\alpha$ being diameter of particle $\alpha$.
The mass and size ratios are $m_L/m_S=2$ and $\sigma_L/\sigma_S = 1.2$, respectively.
$\sigma_S$, $\epsilon/k_B$, and $\tau = (m_S \sigma_S^2/\epsilon)^{1/2}$ are employed as units of length, temperature, and time, respectively.
The number density is set to be $\rho = N/V = (N_L+N_S)/V = 0.8$, and compositions of the two species are the same as $N_L/N = N_S/N = 0.5$.
At $\rho = 0.8$, the MCT power-law fitting for $\tau_{\alpha0}$ estimates $T_c \simeq 0.267$~\cite{Berthier_2012,Kim_2013}.
Note that Refs.~\cite{Berthier_2012,Kim_2013} have studied the case at $\rho = 0.742$ and estimated $T_c = 0.198$.
This value is converted to $T_c \simeq 0.267$ at $\rho = 0.8$ since one dimensionless coupling constant $\Gamma = \rho T^{-1/4}$ determines states of the SS model~\cite{Bernu_1985,Yamamoto_1998}.

Ref.~\cite{Furukawa2_2023} has studied the SS model under shear flow at $T=0.306$, $0.285$, $0.275$, and $0.267$, and reported the viscosity $\eta(T,\dot{\gamma})$ as a function of $\dot{\gamma}$.
Figure~\ref{fig_data} in the main text presents data on $\eta/\eta_0$ versus $\dot{\gamma}/\dot{\gamma}_{c}$ at $T = 0.275$ in (a), and $\dot{\gamma}_c$ versus $\eta_0$ at $T \ge 0.275 > T_c$ in (b).

\subsection*{System description of 2DSS model}
The 2DSS model is a binary mixture composed of large~($L$) and small~($S$) particles in two spatial dimensions~\cite{Yamamoto_1998,Furukawa_2009}.
The particles interact via the inverse power-law potential as described in Eq.~(\ref{potss}).
The mass and size ratios are $m_L/m_S=2$ and $\sigma_L/\sigma_S = 1.4$, respectively.
$\sigma_S$, $\epsilon/k_B$, and $\tau = (m_S \sigma_S^2/\epsilon)^{1/2}$ are employed as units of length, temperature, and time, respectively.
The number density is set to be $\rho = N/V = (N_L+N_S)/V = 0.8$, and compositions of the two species are the same as $N_L/N = N_S/N = 0.5$.
We estimate $T_c \simeq 0.534$ by the MCT power-law fitting on data $\eta_0$ versus $T$ reported in Ref.~\cite{Furukawa_2023}.

Ref.~\cite{Furukawa_2023} has studied the 2DSS model under shear flow at $T=1.43$, $0.85$, $0.665$, $0.577$, and $0.526$, and reported the viscosity $\eta(T,\dot{\gamma})$ as a function of $\dot{\gamma}$.
Figure~\ref{fig_data} in the main text presents data on $\eta/\eta_0$ versus $\dot{\gamma}/\dot{\gamma}_{c}$ at $T = 0.577$ in (a), and $\dot{\gamma}_c$ versus $\eta_0$ at $T \ge 0.577 > T_c$ in (b).

\subsection*{System description of BKS model}
The BKS model is often used for amorphous and supercooled silica~(SiO$_2$)~\cite{Beest_1955}.
Si and O ions interact via the potential,
\begin{equation}
v_{\alpha \beta}(r) = \frac{q_\alpha q_\beta e^2}{r} + A_{\alpha \beta} \exp \left( - B_{\alpha \beta} r \right) - \frac{C_{\alpha \beta}}{r^6},
\end{equation}
where $\alpha$ and $\beta$ denote Si or O.
The values of the partial charges $q_\alpha$ and the constants $A_{\alpha \beta}$, $B_{\alpha \beta}$, and $C_{\alpha \beta}$ are found in Refs.~\cite{Beest_1955,Vollmayr_1996}.
The units of length and time are set to be $2.84\text{\AA}$ and $1.98\times 10^{-13}\text{s}$, respectively.
The temperature is measured in units of $6973.9\text{K}$.
The mass density is fixed at $2.37\text{g/cm}^3$, which corresponds to the number density $\rho = N/V = (N_{\text{Si}}+N_{\text{O}})/V = 1.632$.
The dynamical transition temperature was estimated as $T_c \simeq 3330\text{K} = 0.4775$~\cite{Horbach_1999,Horbach_2001}.

Ref.~\cite{Furukawa_2023} has studied the BKS model under shear flow at $T=0.614$, $0.511$, $0.47$, $0.429$, and $0.39$, and reported the viscosity $\eta(T,\dot{\gamma})$ as a function of $\dot{\gamma}$.
Figure~\ref{fig_data} in the main text presents data on $\eta/\eta_0$ versus $\dot{\gamma}/\dot{\gamma}_{c}$ at $T = 0.511$ in (a), and $\dot{\gamma}_c$ versus $\eta_0$ at $T \ge 0.511 > T_c$ in (b).

\section*{Acknowledgments}
We thank Akira Furukawa and Kang Kim for useful discussions and suggestions.
We also appreciate the referees for their thorough and insightful reviews.
This work was supported by JSPS KAKENHI Grant Numbers 20H00128, 20H01868, 22K03543, 23H04495, and 24H00192.

\section*{Author contributions statement}
H.M. and A.I. designed research; H.M. performed research and analyzed data; H.M. and K.M. wrote the paper; H.M., A.I., T.K., and K.M. discussed the results and commented on the paper.

\section*{Competing Interests}
The authors declare no conflicts of interest.

\section*{Data availability}
The data that support the findings of this study are available within the article.

\bibliographystyle{apsrev4-2}
\bibliography{reference}

\end{document}